# Assembling patchy plasmonic nanoparticles with aggregation-dependent antibacterial activity

Francesco Brasili,[†][a,b] Angela Capocefalo,[†][a] Damiano Palmieri,[a,b] Francesco Capitani,[c] Ester Chiessi,[b] Gaio Paradossi,[b] Federico Bordi [‡][a,d,e] and Fabio Domenici *[‡][b]

[a] Dipartimento di Fisica, Sapienza Università di Roma, Piazzale Aldo Moro 5, 00185 Rome, Italy.

[b] Dipartimento di Scienze e Tecnologie Chimiche, Università di Roma Tor Vergata, Via della Ricerca Scientifica, 00133 Rome, Italy.

[c] Synchrotron SOLEIL, L'Orme des Merisiers, Saint-Aubin, 91192 Gif-sur-Yvette, France.

[d] CNR-ISC UOS Roma, Sapienza Università di Roma, Piazzale Aldo Moro 5, 00185 Rome, Italy.

[e] CNIS Nanotechnology Research Center Applied to Engineering, Sapienza Università di Roma, Piazzale Aldo Moro 5, 00185 Rome, Italy.

**Abstract**

We realise an antibacterial nanomaterial based on the self-limited assembly of patchy plasmonic colloids, obtained by adsorption of lysozyme to gold nanoparticles. The possibility of selecting the size of the assemblies within several hundred nanometres allows for tuning their optical response in a wide range of frequencies from visible to near infrared. We also demonstrate an aggregation-dependent modulation of the catalytic activity, which results in an enhancement of the antibacterial performances for assemblies of the proper size. The gained overall control on structure, optical properties and biological activity of such nanomaterial paves the way for the development of novel antibacterial nanozymes with promising applications in treating multi drug resistant bacteria.

## Introduction

The ability of handling materials at the nanoscale allows for developing multi-functional systems with highly programmable properties for a wide range of applications including biotechnology and nanomedicine [1-3]. Of particular relevance are novel artificial nanomaterials with enzyme-like properties, namely nanozymes, that demonstrated an intrinsic antibacterial activity [4-6] as well as the capability of enhancing or triggering the action of other agents [5]. Such systems are proving to be very effective as self-therapeutic systems in treating multi drug resistant bacteria [7,8].

The antibacterial activity of nanozymes based on inorganic materials typically relies on the release of metal ions or reactive oxygen species [9-11], which interfere with different biological processes including cell metabolism and alter cell membrane stability. For these reasons, these species could have toxic effects also on the cells of the host organism and induce undesired side effects [12-14]. Moreover, it has been recently demonstrated that different bacteria species can develop resistance to silver nanoparticles [15]. These drawbacks can be easily overcome by employing as active component of the nanozyme biomolecules such as proteins, which combine intrinsic biocompatibility with highly selective and specific recognition properties, nearly impossible to achieve using synthetic materials [9].

Among the different scaffolds that could be employed for assembling multi-functional materials, gold nanoparticles (AuNPs) offer several advantages. They are inert and stable under most environmental conditions and exhibit low toxicity [14]. In addition, they allow for easy manipulation and surface conjugation, by both covalent and noncovalent interactions, with functional agents [16-18]. On top of this, AuNPs provide further antibacterial mechanisms, arising from their versatile optical and photothermal properties [13]. These are determined by the excitation of collective electronic oscillations at the metal surface, namely the localised surface plasmons, whose resonance frequency can be tailored by the nanoparticles size, shape and spatial organisation, as well as by the dielectric properties of the surrounding media [18-20]. The efficient conversion of the absorbed light into heat allows for designing photothermal vectors capable to burst bacteria [21]. Moreover, the possibility of tuning the optical properties in a wide range of frequencies, spreading from visible to infrared, opens for the *in vivo* application of these plasmonic devices, taking advantage of the transparency of blood and tissues in near infrared (NIR) [22]. A key issue in designing versatile nanoparticle-based materials is to reach a strict control on the assembly process underlying the spatial arrangement of AuNPs. An effective strategy adopted for controlling the organisation in solution relies on interfacing AuNPs with biomolecules, whose programmable intermolecular interactions provide the opportunity of assembling hybrid systems with the desired structure and functionality [23-25]. In addition, their responsiveness to external stimuli such as temperature, pH and incident light supplies further degrees of freedom in controlling the properties and the assembly of the whole system [26]. Thus, in order to take full advantage of such potentialities, it is essential to gain a fine control not only on the guided assembly, but also on the effects prompted by the environmental conditions. In this respect, several studies report on the protein-induced aggregation of AuNPs [27-30], but a comprehensive explanation of all the mechanisms involved is still lacking. In particular, how the adsorption of molecules affects the colloids surface properties and triggers the aggregates formation is far from being fully elucidated.

In this work, we developed a novel plasmonic nanozyme with tunable antibacterial properties based on the assembly of patchy AuNPs. Colloids with inhomogeneous surface charge were obtained by adsorption of lysozyme to anionic AuNPs. The arising charge-patch interactions allow for driving their self-limited aggregation into stable clusters with selected finite size [31,32]. We chose lysozyme (Lyz), an antimicrobial enzyme with size of 3 nm, for its stable globular folding and positive net charge at physiological pH due to the high isoelectric point at pH 11.3.

We performed a thorough analysis of the assembly of Lyz-decorated AuNPs (Lyz-AuNPs) as a function of all the experimental parameters involved in the process to gain a close control on the fabrication of Lyz-AuNPs assemblies. We therefore focused on the antibacterial function of the system, in terms of both plasmonic and catalytic properties, in relation to the colloidal assembly with the aim of highlighting their strict integration and interplay and realising a nanozyme with a high level of tunability.

## Materials and methods

### Materials

Citrate-stabilised AuNPs with nominal size of 100 nm and 60 nm were provided by Ted Pella Inc. The concentrations of the stock solutions were 9.3 pM and 43 pM, respectively. Chicken egg white lysozyme powder (purity ≥ 90%) and (3-Aminopropyl)triethoxysilane (APTES, purity ≥ 98.5%) were provided by Sigma-Aldrich. Sodium citrate buffers at pH 6.5 and pH 4.0 were provided by Merk Millipore. The components of the system were characterised by Dynamic Light Scattering (DLS) measurements, in terms of ζ-potential and hydrodynamic diameter distributions, and UV-Visible absorption spectroscopy in the pH conditions employed for the experiments. The centre values of ζ-potential and hydrodynamic diameter ($2R_H$) distributions are reported in Table 1 together with the centre wavelength of the Localised Surface Plasmon Resonance (LSPR) of AuNPs. For the characterisation, lysozyme stock solutions were prepared at the concentration of 70 μM.

### Sample preparation

Samples were prepared at room temperature. At first, lysozyme was dissolved in 20 mM sodium citrate buffer to obtain solutions with different concentrations at controlled pH. Lyz-AuNPs colloids were therefore prepared at different protein-AuNP molar ratios $x$ by adding to each protein solution the same volume of AuNPs stock solution and incubating for 2 minutes at room temperature. The final concentration of citrate buffer is therefore of 10 mM in all samples.

Preliminary control experiments have been performed by UV-Visible absorption spectroscopy to determine the optimal citrate buffer concentration that maintains the pH at the selected value without altering the AuNPs capping (i.e., without inducing shifts in the LSPR).

Measurements as a function of pH were performed according to the following protocol. Lyz-AuNPs solutions were prepared at pH 4. The pH of each solution was then changed to pH 6.5 by adding NaOH. The pH was brought back to 4 by adding HCl.

**Table 1.** DLS and UV-Visible characterisation of AuNPs and of lysozyme in the two pH conditions employed for the experiments. Each reported value is obtained by three measurements.

| System component | pH | ζ-potential (mV) | $2R_H$ (nm) | resonance wavelength (nm) |
|---|---|---|---|---|
| AuNPs 100 nm | 6.5 | -54 ± 2 | 102 ± 2 | 572.2 ± 0.1 |
| AuNPs 60 nm | 6.5 | -42 ± 3 | 63 ± 2 | 536.3 ± 0.1 |
| Lysozyme | 6.5 | +4.0 ± 0.3 | 2.9 ± 0.2 | -- |
| AuNPs 100 nm | 4.0 | -48 ± 2 | 102 ± 2 | 572.2 ± 0.1 |
| Lysozyme | 4.0 | +14 ± 1 | 2.9 ± 0.2 | -- |

### Dynamic Light Scattering

For DLS measurements we employed a NanoZetaSizer apparatus (Malvern Instruments LTD), equipped with a 5 mW He-Ne laser (633 nm wavelength). Experiments were performed at 25°C. To obtain the intensity weighted distributions of the hydrodynamic diameter (i.e., the diameter of a sphere with the same diffusion coefficient of the particle), the intensity autocorrelation functions were acquired at an angle of 173°. Correlograms were analysed with the CONTIN algorithm to extrapolate the associated decay times (Section S1 of ESI). Decay times are used to determine the distribution of the diffusion coefficients D of the particles, which in turn are converted in a distribution of hydrodynamic diameters $2R_H$ using the Stokes-Einstein relationship $R_H = k_BT/6\pi\eta D$, where $k_BT$ is the thermal energy and $\eta$ the solvent viscosity.

The ζ-potential was obtained by combining laser Doppler velocimetry and phase analysis light scattering to accurately determine the average electrophoretic mobility $\mu_e$ and its distribution. The measured values were converted into the ζ-potential using the Smoluchowski relation:

$$\zeta = \frac{\mu_e \eta}{\varepsilon} \qquad (1)$$

where $\varepsilon$ is the solvent permittivity.

The data analyses performed to extrapolate hydrodynamic diameter and ζ-potential distributions were performed using the Zetasizer software provided together with the instrument. The reported values are obtained by averaging the centre values of at least three distributions measured independently. The associated error is the corresponding standard deviation.

### Absorption Spectroscopy

For UV-Visible-NIR absorption spectroscopy we employed a v-570 double ray spectrophotometer (Jasco), with a resolution of 0.1 nm in the UV-Visible region and 0.5 nm in the NIR region. The spectrophotometer is equipped with a Peltier thermostat ETC-505T (Jasco) to keep the sample temperature at 25°C. The reported extinction spectra are normalised to the absorbance at 400 nm. At this wavelength the absorption coefficient is proportional to the molar concentration of Au(0) in the sample and is assumed to be independent on the AuNPs size [30].

Spectra analysis was performed by a band fitting procedure with two Gaussian curves to disentangle the contribution to each extinction spectrum of the LSPR of single AuNPs from that of the inter-particle plasmonic modes due to AuNPs aggregation.

The LSPR wavelength of the non-aggregated Lyz-AuNPs was obtained by fitting the resonance to a Gaussian line shape to determine the peak position. Each LSPR wavelength value is the average on three independent measurements and the associated error is the corresponding standard deviation.

All the analyses were performed with Origin 8.1 software.

### Atomic Force Microscopy and Scanning Electron Microscopy

Samples for near-field microscopy imaging were prepared at room temperature, by incubating 50 μL of sample solution for 10 minutes onto a silicon substrate previously functionalised with APTES. The deposition procedure based on the substrate derivatisation was chosen and accurately optimised with respect to deposition time in order to minimise possible modification of the aggregates structure during adhesion to the silicon surface [33].

Atomic Force Microscopy (AFM) images were recorded using a Dimension Icon Bruker microscope in tapping mode, with a scan rate of 0.5 Hz. A cantilever with a spring constant of 42 N/m and a tip with a nominal radius of curvature of 2 nm was employed. AFM images were analysed by Gwyddion software, version 2.52.
Scanning Electron Microscopy (SEM) images were recorded using a Zeiss Auriga 405 Field Emission Scanning Electron Microscope. For the evaluation of the inter-particle distances in the deposited Lyz-AuNPs clusters, whose details are reported in ESI, SEM images have been analysed by ImageJ software, version 1.48v.

**Synchrotron Radiation FTIR microspectroscopy**

Synchrotron Radiation FTIR microspectroscopy (SR-microFTIR) was performed at the SMIS beamline of SOLEIL Synchrotron facility (Saint-Aubin, France). Lyz-AuNPs samples for SR-microFTIR measurements were separated from non-adsorbed lysozyme by centrifugation, deposited by drop-casting procedure onto a double-polished silicon substrate and dried at room temperature.

IR spectra of selected areas were measured in transmission with a Continuum XL (Thermo Fisher Scientific) IR microscope equipped with a liquid nitrogen cooled MCT detector, a 32× Schwarzschild objective, a motorized aperture and stage. The microscope was coupled to a Nicolet 5700 FTIR spectrometer (Thermo Fisher Scientific).

We used an aperture of 8 μm × 8 μm, a spectral resolution of 4 cm$^{-1}$ and 200 scans for each acquisition. A background spectrum was collected for every batch of spectra through a clean area of the silicon substrate. Spectra reported in the present paper are the average of at least 5 acquisitions and are presented after baseline subtraction, smoothing procedure by the Savitzky-Golay algorithm (2$^{nd}$ order polynomial, 8 data points) and normalisation to the maximum absorbance value. Spectral analysis was performed using Origin 8.1 software.

**Antibacterial activity assay**

The antibacterial activity of samples was assayed at pH 6.5 and 25°C by employing a commercial kit (Sigma Aldrich). A cell suspension was prepared by dissolving lyophilised *Micrococcus lysodeikticus* bacteria in a potassium phosphate buffer at 0.01 % (w/v). For each experiment, 30 μL of the sample were added to 800 μL of bacteria suspension. A turbidimetric assay was performed to evaluate the lytic activity $\varrho$ by recording the decrease in time of the absorbance at 450 nm $A_{450}$ with the Jasco v-570 double ray spectrophotometer. Typically, kinetic runs were followed for 15 minutes. The activity rate was determined by the slope $dA_{450}/dt$ of the initial linear region of the absorbance trend, according to the formula [34]:

$$\varrho = \frac{\frac{dA_{450}}{dt}\big|_{sample} - \frac{dA_{450}}{dt}\big|_{blank}}{\frac{dA_{450}}{dt}\big|_{unit}} \quad (2)$$

where $\frac{dA_{450}}{dt}\big|_{unit} = 3\times10^{-5}$ abs. unit/min is the slope induced in the absorbance at 450 nm by one unit of active lysozyme, rescaled to the actual volume employed in the experiments (30 μL).

To evaluate the lysis efficiency of the proteins confined in AuNPs clusters, the samples were centrifuged to separate non-adsorbed protein (supernatant) from the Lyz-AuNPs complexes. We then performed independent turbidimetric assays on the Lyz-AuNPs overall sample, on the supernatant and on the free protein at the same concentration employed to prepare the sample. We therefore calculated the normalised activity $\varrho_{norm}$ of the lysozyme confined within clusters by the formula:

$$\varrho_{norm} = \frac{\varrho_{Lyz-AuNPs} - \varrho_{supernatant}}{\varrho_{free\,protein} - \varrho_{supernatant}} \quad (3)$$

where $\varrho_{Lyz-AuNPs}$, $\varrho_{supernatant}$ and $\varrho_{free\,protein}$ are the activities of the Lyz-AuNPs overall sample, of the non-adsorbed proteins and of the free protein, respectively.

The reported values are obtained by averaging at least three measurements. The associated error is the corresponding standard deviation. Data analysis was performed with Origin 8.1 software.

## Theoretical background

It is well known (see for example ref. [32] and the literature cited therein) that a non-uniformly distributed electric charge at the surface of colloidal particles in aqueous solution results in an inter-particle potential that, even if the net charges on the two particles have the same sign, may show a significant attractive component. Intuitively such attraction originates from the interplay of short range, local interactions between oppositely charged patches on the approaching particles and the overall screening due to the ionic strength of the solution. In this section, we briefly recall the theoretical model for charge-patch interactions developed by Velegol and Thwar [31], and we derive an analytical expression for the standard deviation $\sigma$ of the surface potential that accounts for the surface charge inhomogeneity.

**Interaction potential between charge-patched colloids**

The theory of Velegol and Thwar for charge-patch interactions is based on the Derjaguin approximation [31]. In this approximation, the interaction force $F(h)$ between two spherical particles can be expressed as a function of the separation $h$ between their surfaces as:

$$F(h) \propto \frac{a_1 a_2}{a_1 + a_2} W(h) \quad (4)$$

where $a_1$ and $a_2$ are the two radii of curvature of the particles, and $W(h)$ is the interaction potential between two flat surfaces at the same distance. This expression highlights that the particles sizes contribute only through a scaling factor and do not affect the functional form.

To calculate the pair interaction potential $V(h)$, which reduces to $W(h)$ for flat surfaces (*i.e.*, for infinite radii of curvature $a_1, a_2 \to \infty$), the theory assumes a random distribution of the charge patches on the colloids surfaces. In the case of identical particles ($a_1 = a_2 = a$), $V(h)$ can be expressed in terms of their average surface potential $\zeta$ and of the corresponding standard deviation $\sigma$:

$$V(h) = \pi\varepsilon a \left[(\zeta^2 + \sigma^2)\ln(1 - e^{-2\kappa h}) + \zeta^2 \ln\left(\coth\frac{1}{2}\kappa h\right)\right] \quad (5)$$

where $\varepsilon$ is the permittivity of the dispersing medium and $\kappa^{-1}$ is the Debye screening length. The interaction potential combines two terms: a repulsive monopole and an attractive multipole, which depend on $\zeta$ and $\sigma$, respectively [32]. Having different interaction ranges, the two terms give rise to a potential barrier that the particles must overcome to stick together, whose height $V_{max}$ is given by:

$$V_{max} = \pi\varepsilon a \left\{(\zeta^2 + \sigma^2)\ln\left[1 - \left(\frac{\zeta^2}{\zeta^2+\sigma^2}\right)^2\right] + \zeta^2 \ln\left(\frac{2\zeta^2+\sigma^2}{\sigma^2}\right)\right\} \quad (6)$$

In our case, the spherical particles are the Lyz-AuNPs, whose average surface potential has been evaluated by ζ-potential measurements. The non-uniformity of the surface charge distribution is due to the adsorption of lysozyme, which sparsely decorates the colloids surface. If the net charge carried by each protein is higher than that of the portion of the AuNP surface on which the molecule adsorbs, the resulting patches bear a net charge which is opposite in sign to that of the bare particle surface.

**Standard deviation of the surface potential in charge-patched colloids**

The standard deviation of the surface potential is defined as:

$$\sigma = \sqrt{(\zeta_0 - \zeta)^2(1-\Phi) + (\zeta_{cov} - \zeta)^2 \Phi} \qquad (7)$$

where $\Phi$ is the surface coverage of the colloids (*i.e.*, the portion of the AuNPs surface covered by the adsorbed proteins), while $\zeta_0$ and $\zeta_{cov}$ are the values of the surface potential in correspondence of the non-covered and covered portions of the particles surface, respectively. Thus, assuming that the surface charge density of the non-covered regions is not affected by the protein adsorption, $\zeta_0$ can be evaluated by the ζ-potential of the bare colloids and the variation $\Delta\zeta$ of the average surface potential is proportional to $\Phi$:

$$\Delta\zeta = \zeta - \zeta_0 = (\zeta_{cov} - \zeta_0)\Phi \qquad (8)$$

Combining equation 7 and 8 and defining $\Delta\zeta_{max} = \zeta_{cov} - \zeta_0$ the two equivalent analytical expressions for $\sigma$ are obtained:

$$\sigma = \Delta\zeta_{max}\sqrt{\Phi(1-\Phi)} = (\zeta - \zeta_0)\sqrt{\frac{1-\Phi}{\Phi}} \qquad (9)$$

The first expression is equivalent to that previously reported [35], as discussed in Section S2 of ESI. It identifies two distinct contributions to the surface charge inhomogeneity: an electrical one, $\Delta\zeta_{max}$, which is determined by the net charge transported by each adsorbate molecule and a geometrical one, arising from $\Phi$, which is the product of the total number of adsorbed molecules by the area ratio between the fingerprint of a single adsorbed molecule and the total surface of the colloidal particle. The second expression allows for calculating $\sigma$ by combining the measured ζ-potential values with an estimate of the surface coverage as discussed in ESI.

## Results and discussion

The assembly of citrate-stabilised AuNPs upon mixing with a lysozyme solution was studied with the aim of gaining a fine control on the different features (optical response and catalytic activity) involved in the antibacterial activity of the resulting system. A scheme of the assembling strategy adopted and of the investigation performed is reported in Figure 1. The aggregation in solution is the pivotal mechanism, thus a detailed characterisation of the process was performed as a function of the different parameters involved, in the framework of the charge-patch interactions. Such phenomenology, whose theory is briefly recalled in Theoretical background, is characterised by a self-limited assembly into fractal-like clusters with tunable size. DLS, near field microscopies and absorption spectroscopy measurements were combined in order to directly relate the plasmonic response of the Lyz-AuNPs complexes to their structural and electrostatic features. The activity rate of the complexes in bursting bacteria cells was assayed in order to identify the experimental parameters that provide a more efficient antibacterial action.

**Charge-patch interactions towards a controlled aggregation**

The colloidal aggregation of AuNPs upon adsorption of lysozyme was studied at varying the different experimental parameters affecting the inhomogeneity $\sigma$ of the surface potential, which gives rise to the attractive component of the interaction potential of equation 5. Focusing on the geometrical contributions to $\sigma$ (see equation 9), we varied the AuNPs surface coverage $\Phi$ by acting on the number of adsorbed molecules through the molar ratio $x$ between proteins and colloids, and on the area ratio between the fingerprint of a single adsorbed lysozyme molecule and the total surface of each colloid by using two sizes for AuNPs, 100 nm and 60 nm. It is worth nothing that the colloids size also affects the scaling factor of the interaction potential. In both cases the AuNPs size is one order of magnitude higher as compared to the protein allowing for finely tuning the charge inhomogeneity on the colloids surface.

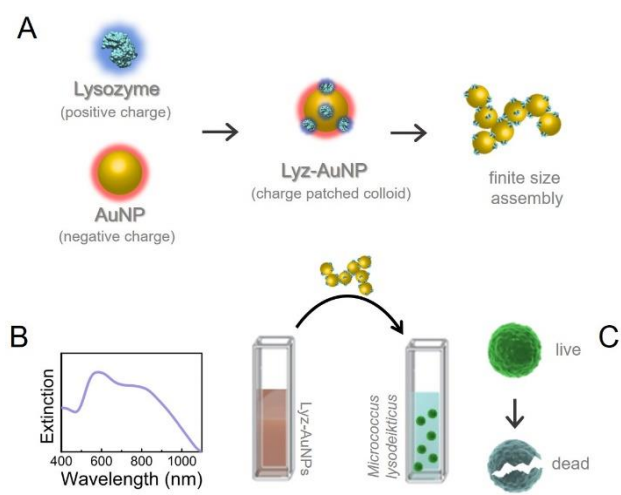

**Figure 1.** Schetches illustrating the strategy adopted for synthesising Lyz-AuNPs and the experimental plan. (A) Formation of charge-patched colloids by electrostatic adsorption of positively charged proteins onto anionic AuNPs and consequen aggregation into clusters with finite size. (B) Optical characterisation of the solution containing Lyz-AuNPs complexes. (C) Assay of the catalytic activity of complexes performed on the gram-positive *Micrococcus lysodeikticus* bacteria.

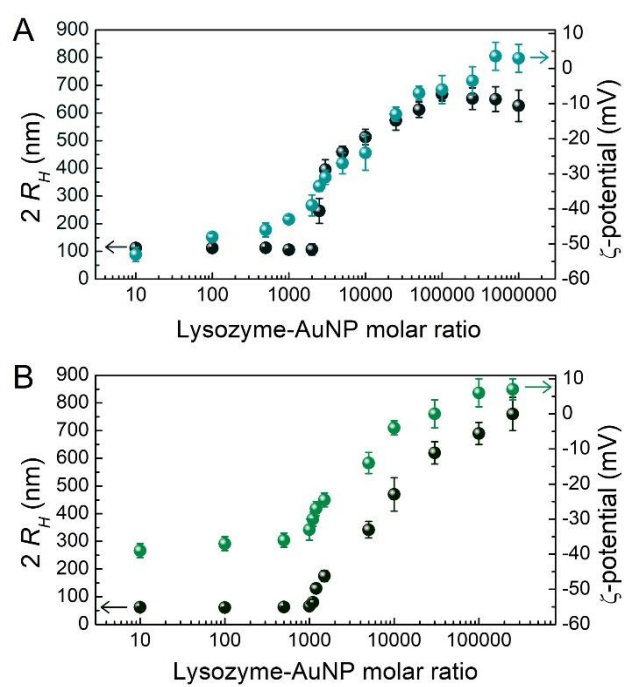

**Figure 2.** DLS measurements (hydrodynamic diameter, $2R_H$, and ζ-potential trends) on Lyz-AuNPs as a function of the lysozyme-AuNPs molar ratio, using AuNPs of 100 nm (A) and 60 nm (B). Each reported value is obtained by three measurements.

The hydrodynamic diameter and ζ-potential trends of Lyz-AuNPs as a function of $x$ are reported in Figure 2. The experiments point out a self-limited aggregation of colloids into assemblies whose size is determined by the lysozyme-AuNP molar ratio, consistently with a charge-patch driven aggregation [35,36]. This type of process is characterised by the formation of fractal assemblies [37,38]. Hydrodynamic diameter is commonly employed as reliable indicator of size evolution trends for fractal systems [39,40], nevertheless it should be taken into account that the measured values could result in an underestimation of the actual maximum extension of such objects. The cluster stability has been evaluated by the steadiness of the hydrodynamic diameter distributions after repeated measurements (see Section S1 of ESI).

Three different stages of the aggregation phenomenology can be recognised. At low molar ratios ($x \leq 2000$ for 100 nm AuNPs, panel A, and $x \leq 1000$ for 60 nm AuNPs, panel B) hydrodynamic diameter values remain stable around the size of single AuNPs, while slight variations occur in the ζ-potential of colloids. This points out that few proteins adsorb to AuNPs and consequently the long-range electrostatic repulsion remains the predominant component of the interaction between AuNPs (equation 5), preventing aggregation.

For higher molar ratios, starting from the threshold values of $x_{100} \approx 2500 \pm 500$ for 100 nm AuNPs and $x_{60} \approx 1100 \pm 100$ for 60 nm AuNPs, a steeper variation of the ζ-potential is observed, concomitant to the formation of aggregates. In this stage, the surface charge inhomogeneity originated by adsorbed lysozyme molecules induces a significant short-range attraction. Note that the amount of lysozyme adsorbed to AuNPs is only a portion of that one present in solution. In fact, the aggregation process into clusters of growing size proceeds even for molar ratios extremely higher than the number $N_{max}$ of proteins corresponding to the full coverage of the AuNPs surface (which can be evaluated by geometrical considerations to ~ 4900 for 100 nm AuNPs and ~ 1800 for 60 nm AuNPs [41]).

When the ζ-potential values approach zero, the hydrodynamic diameter distributions show the tendency to become broader and unstable. Correspondingly, aggregates reach the maximum values of the hydrodynamic diameter of 640 ± 30 nm for 100 nm AuNPs and 760 ± 60 nm for 60 nm AuNPs.

For the highest molar ratios (starting from $x \approx 500000$ for 100 nm AuNPs and from $x \approx 100000$ for 60 nm AuNPs) a slight overcharging, that is the inversion of the complexes charge sign with respect to that of the bare colloids, is also observed, with positive ζ-potential values always lower than 10 mV. In this range, both the hydrodynamic diameter and ζ-potential distributions do not evolve anymore at increasing the molar ratio. This could be explained by the hindering of the adsorption of more lysozyme molecules onto the AuNPs due to both steric hindrance and/or electrostatic repulsion between the approaching lysozyme molecules and those already adsorbed.

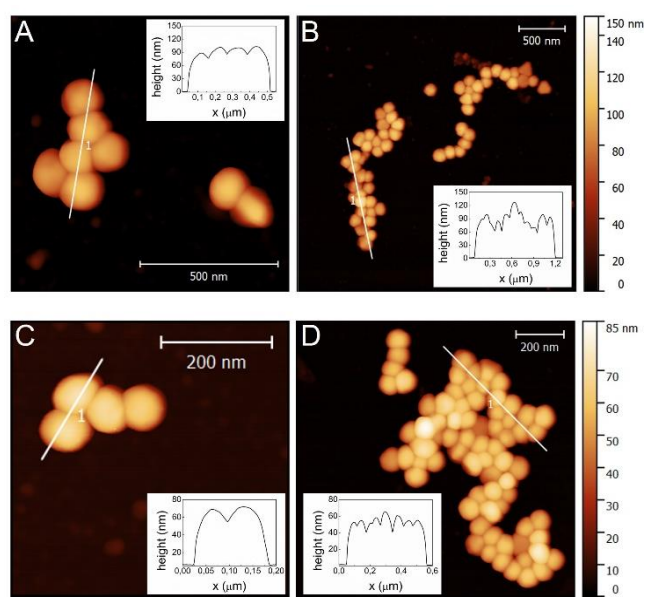

**Figure 3.** Representative AFM images acquired on clusters corresponding to $x = 3000$ (A) and $x = 10000$ (B) for 100 nm AuNPs and to $x = 1200$ (C) and $x = 5000$ (D) for 60 nm AuNP. The corresponding height profiles are reported in the insets.

Comparing the two AuNPs sizes investigated, it can be noticed that the onset of the aggregation occurs for similar values of the potential barrier $V_{max}$. In fact, the threshold ζ-potential values are ζ = -33.5 ± 1.5 mV for 100 nm AuNPs and ζ = -30 ± 2 mV for 60 nm AuNPs. The repulsive contribution to the interaction energy between colloids is therefore of the same order of magnitude. The attractive contribution is also similar in the two cases. In fact, the ratio $x_{100}/x_{60} = 2.3 \pm 0.5$ is consistent with the ratio between the surface areas of the two colloids (equal to 2.8). Assuming that the number of adsorbed molecules only depends on the protein concentration in solution (i.e., the adsorption of a molecule is not affected by the presence of other molecules onto the gold surface), the inhomogeneity degree on the colloids surfaces (i.e., the portion of the surface covered with proteins) is roughly the same at the aggregation onset. A second point regards the maximum hydrodynamic diameter, which is slightly larger in the case of the smaller AuNPs. This finding is consistent with the theoretical models (see equations 4 and 6), noting that the limiting size above which two approaching clusters do not aggregate anymore is reached faster when the primary particles are larger.

A quantitative analysis of the aggregation phenomenology can be carried out from the ζ-potential values measured on the non-aggregated Lyz-AuNPs, as discussed in Section S2 of ESI. We derived the number $N_{lyz}$ of proteins adsorbed to each AuNP, the surface coverage $\Phi$, the standard deviation $\sigma$ of the surface ζ-potential, and the height $V_{max}$ of the potential barrier. Notably, for both AuNPs sizes, at the onset of the aggregation process the surface coverage is lower than 5%, while the calculated values of $V_{max}$ approach the thermal energy (~ 25 meV at 25°C), as highlighted in Figure S3 of ESI.

To gain more insight on the morphology of the Lyz-AuNPs, AFM measurements were performed on clusters deposited and dried on silicon substrates. Representative images are reported in Figure 3. Consistently with the DLS results, the size of the aggregates measured by AFM increases with the lysozyme-AuNP molar ratio $x$. Clusters with low coordination number (Figures 3A and 3C), observed just above the aggregation threshold, appear compact, while at higher molar ratios (Figure 3B and 3D) they became less regular and more branched. In this fractal-like organisation, resulting upon the two-dimensional rearrangement, the aggregates are mainly composed by one stack of AuNPs (superimposed AuNPs are rarely observed), suggesting a branched and sparse three-dimensional conformation. Consistently, preliminary Small Angle X-ray Scattering (SAXS) analysis reported in Section S3 of ESI highlights the low fractal dimension, always lower than 2, of the dispersed clusters.

To further assess charge-patch interactions as the driving mechanism for the aggregation phenomenology, we studied the process in a solution with pH 4, focusing on 100 nm AuNPs. In such conditions we would expect variations in the aggregation behaviour, due to the pH-dependent charge of both proteins and AuNPs. Below the isoelectric point, in fact, the overall protein charge shows a clear dependence on the pH [42], becoming markedly positive in acidic environment. Specifically, the measured $\zeta$-potential shifts from 4 mV at pH 6.5 to 14 mV at pH 4 (see Table 1). With reference to equation 9, this directly affects $\Delta\zeta_{max}$, and therefore allows for inducing high inhomogeneity in the surface potential already at low surface coverage, that is at low molar ratios. In addition, the protonation of some carboxylic groups of the citrate molecules capping the AuNPs, occurring at pH 4, induces a slight shift in the $\zeta$-potential of the bare 100 nm AuNPs. We measured a $\zeta$-potential of -48 mV at pH 4 and of -54 mV at pH 6.5 (Table 1). This lowers the contribution of the repulsive component of the interaction potential of equation 5.

The $\zeta$-potential and hydrodynamic diameter as a function of the lysozyme-AuNP molar ratio measured at pH 4 are compared with the corresponding quantities measured at pH 6.5 in Figure S9 of ESI. While the overall experimental trends appear similar for both pH conditions, at the lower pH AuNPs aggregation is triggered at lower lysozyme concentrations ($x \approx 500$), and at the same molar ratio larger clusters are observed. The calculated height $V_{max}$ of the potential barrier at the onset of the aggregation is of the same order of magnitude of the thermal energy $k_BT$, as highlighted in Figure S10. The isoelectric point of the complexes is reached at $x = 10000$, one order of magnitude lower than that at pH 6.5. Correspondingly, the maximum size of the Lyz-AuNPs is of ~ 700 nm. For higher lysozyme concentrations ($x \geq 25000$), a marked overcharging is observed, and the formed aggregates have a reduced size. This phenomenon, known as re-entrant condensation [32], occurs when the amount of proteins adsorbed onto the colloids surface is higher than that needed to completely neutralise the $\zeta$-potential, inducing a significant overcharging. The dependence of the aggregation process on the pH of the solution, and thus on the net charge of the system components, clearly assesses the role of the electrostatic interactions in the cluster formation and stability, and confirms that acting on their strength it is possible to control the cluster formation.

**Tailoring the plasmonic response of the nano-assemblies**

In the previous Section we demonstrated that the protein-mediated assembly of AuNPs represents an efficient route for the fabrication of clusters with controllable finite size. Proceeding from this, we investigated in detail the optical response of the obtained aggregates aiming at tailoring their plasmonic response (i.e., the LSPR), which is directly involved in the intrinsic antibacterial function of AuNPs [13,21]. To this end, we measured the extinction spectra of Lyz-AuNPs complexes assembled at selected values of the molar ratio

$x$, according to the aggregation trends of Figure 2. Representative spectra are reported in Figure 4 for 100 nm (panel A) and 60 nm (panel B) AuNPs.

For both the AuNP sizes the spectra show similar evolution, depending on the Lyz-AuNPs molar ratio. At low $x$ values, the peak corresponding to the LSPR of AuNPs shows a progressive intensity quenching and a redshift compared to that of the stock solution. Starting from $x \approx 2500$ for 100 nm AuNPs and from $x \approx 1100$ for 60 nm AuNPs, a shoulder at higher wavelengths appears in the plasmonic profiles, together with a further broadening and shift of the LSPR of the primary colloids. The spectral changes proceed at increasing the molar ratio with the rise of a wider band well-extended into the NIR region, more extensively for 100 nm AuNPs, up to an asymptotic condition (starting from $x \approx 50000$ for 100 nm AuNPs and from $x \approx 10000$ for 60 nm AuNPs).

A strict correspondence can be recognised between the evolution of the Lyz-AuNPs extinction spectra and the stages of the colloidal aggregation discussed above. The initial redshift and quenching of the LSPR peak are consistent with changes in the dielectric environment at the AuNPs interface induced by the adsorption of lysozyme [43]. This is in accordance with the slight increase of the measured $\zeta$-potential. The shoulder in the plasmonic profiles, which appears in correspondence of the onset of the aggregation process, originates from the rise of coupled inter-particle plasmonic modes due to the constructive interference between the electronic oscillation modes of the single AuNPs [44]. The redshift and broadening of the extinction band observed upon increasing the lysozyme concentration reflect the hydrodynamic diameter and $\zeta$-potential increasing trends of Figure 2. Analogously, further

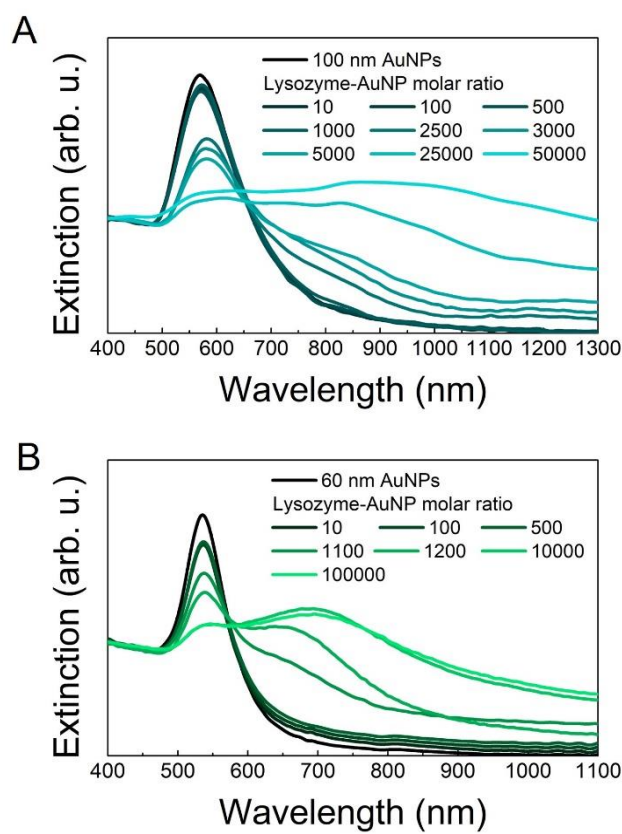

**Figure 4.** Representative extinction spectra acquired at different Lyz-AuNP molar ratios for 100 nm (A) and 60 nm (B) AuNPs. The spectra measured for the corresponding AuNPs stock solutions are also shown for comparison.

evolution of the optical response is not observed after the isoelectric point of the complexes.

The modifications in the extinction spectra correlated to the presence of a specific molecule in solution have been widely exploited in the literature for developing ultrasensitive colorimetric assays [45]. In this framework, we established in Section S4 of ESI a quantitative correspondence between the spectra and the aggregation phenomenology by defining a parameter given by the ratio between the area underlying the band of the inter-particle plasmonic modes and the total area of each extinction spectrum. The trends as a function of the Lyz-AuNPs molar ratio obtained using the normalised area show a strict accordance with those of Figure 2. This result demonstrates that the normalised area represents a valuable experimental parameter allowing to follow the aggregation of Lyz-AuNPs directly from the extinction spectra.

By analysing the initial redshift of the plasmonic profiles, highlighted in Figure S6 of ESI, it is possible to extrapolate quantitative information on the lysozyme adsorption to AuNPs. Exploiting the dependence of the resonance wavelength on the effective refractive index at the nanoparticle interface, we derived an expression for the redshift induced in the LSPR by a non-uniform adsorbate layer, as a function of the surface coverage $\Phi$. The detailed calculation, reported in Section S5 of ESI, proceeds from the quasi-static approximation for the localised electric field [46] and modifies the expression for the polarizability of AuNPs [47] to account for retardation effects. The values of $\Phi$, calculated for each sample, are reported as a function of the lysozyme concentration in the isotherm curves of Figure S7. Such values are consistent with those extrapolated from the ζ-potential measurements (see Tables S1 and S2). On the basis of these results, non-aggregated AuNPs emerge as an appealing sensor able to detect small amounts of protein molecules in solution, down to pM concentrations (3.7 pM for 100 nm AuNPs and 15.5 pM for 60 nm AuNPs). Moreover, a fit of the experimental trend by Langmuir isotherm model yields the apparent dissociation constant $K_d$ for the protein adsorption to AuNPs. The values obtained are $K_d = 13 \pm 3$ pM for 100 nm AuNPs and $K_d = 80 \pm 20$ pM for 60 nm AuNPs. These values are consistent with those reported in literature for the adsorption of proteins to nanoparticles [48-50]. The lower $K_d$ obtained for 100 nm AuNPs (∼ 6-fold) can be ascribed to the ζ-potential, which is larger in modulus as compared to 60 nm AuNPs (Table 1). The larger radius of curvature may also contribute to strengthen the binding of adsorbing proteins, entailing that a wider portion of the nanoparticle surface is involved in the interaction.

Focusing on the spectra corresponding to the aggregation onset, the observed changes are ascribable to the formation of dimers and small aggregates [51,52]. The further redshift and broadening of the band at increasing the lysozyme amount point out a higher number of coupled plasmons due to the formation of larger aggregates, which absorb light in a wider range of wavelengths. Such optical behaviour is consistent with the plasmonic response of three-dimensional fractal aggregates of AuNPs, characterised by the superposition of several inter-particle modes corresponding to different interaction strengths between plasmons [52,53]. To better investigate this aspect, we estimated the inter-particle distance distributions on SEM images of samples deposited and dried on silicon substrates using a protocol accurately optimised for minimising structural alterations of complexes during adhesion. The analysis of these images, reported in Section S7 of ESI, yielded separations of few nanometres, increasing for larger and more branched aggregates. This finding is consistent with the analysis of the SAXS structure factors reported in Table S3 of ESI. The wide, inhomogeneous distributions of the inter-particle distance for both the AuNPs sizes justify the observed non-sharp separation between longitudinal and transverse plasmonic modes. Actually, it would be interesting to analyse in detail this correspondence between structure and optical response of clusters by near-field electromagnetic simulations of the coupled plasmonic modes [52,53], for which the microscopic and spectroscopic characterisation herein reported would represent an important experimental basis.

The systematic study of the optical response of Lyz-AuNPs was conducted also on samples prepared at pH 4. Selected extinction spectra are reported in Figure S11 of ESI. Also in this case the optical response of the samples strictly follows the aggregation trend (see Figure S12). Interestingly, it is even possible to recognise the re-entrant condensation from the extinction spectra, which correspondingly show a shrinkage and a blueshift of the inter-particle plasmonic band together with an increase of the peak corresponding to the plasmonic modes of the primary colloids.

From the above analysis, a clear correspondence between the optical behaviour of the system and the size and the morphology of the Lyz-AuNPs clusters emerges, unveiling the potentiality of our route to synthesise in solution nanomaterials with the desired optical properties. Noteworthy, especially in the case of the larger AuNPs, the aggregation bands are very prominent in the NIR spectral region, making the system intriguing for *in vivo* applications [21,22], due to the reduced absorption of biological tissues in this wavelength range. On this line, our further investigation was focused on the 100 nm AuNPs.

By exploiting the dependence of the aggregation process on the pH of the solution, we explored the possibility of employing this parameter to modulate the optical and structural properties of pre-assembled clusters. A representative example is shown in Figure 5, where the extinction spectra (panel A) of the sample prepared at $x = 2000$ are shown together with the corresponding hydrodynamic diameter (panel B) and ζ-potential (panel C) distributions. The images enlighten modification in both the optical response and the aggregation process, occurring when the pH of the solution is changed from 4 to 6.5 and back to 4. Specifically, when the pH increases the net charge transported by each protein lowers and the resulting surface charge distribution on the lysozyme-decorated AuNPs (ζ-potential shifted from -20 mV to -40 mV) can no longer maintain the AuNPs aggregated in the initial configuration. Their disaggregation is therefore observed with the mean hydrodynamic diameter of the aggregates decreased from 500 nm to 150 nm and the extinction spectrum showing the narrowing and blueshift of the plasmonic inter-particle band together with an intensity increase of the single particle LSPR. When the pH is lowered back, an opposite behaviour is observed, resulting in a re-aggregation of the Lyz-AuNP complexes, as proved by the increase of the hydrodynamic diameter and the broadening of the inter-particle plasmonic band (light red curves), pointing out the reversibility of the phenomenon. The slight differences observed between the ζ-potential values measured at pH 4 before and after the modulation can be ascribed to the increased ionic strength in the solution, resulting from the addition of NaOH and HCl, and the consequent decrease of the effective range of the screened electrostatic forces that increases the importance of the attractive contribution of van der Waals interactions [32, 35].

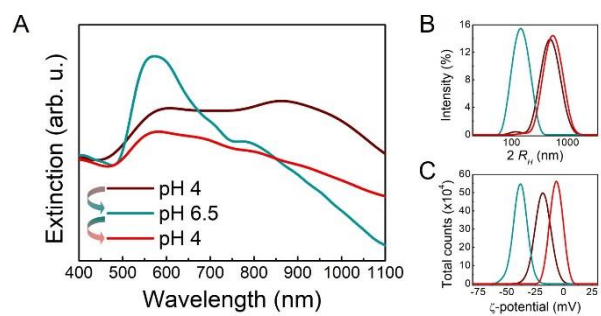

**Figure 5.** Modulation of the Lyz-AuNPs assembling and optical properties by acting on the pH of the solution: representative extinction spectra (A), hydrodynamic diameter (B) and ζ-potential (C) distributions of 100 nm Lyz-AuNPs prepared at $x = 2000$.

**Assaying the antibacterial activity on gram-positive bacteria**

Lysozyme is an antimicrobial enzyme catalysing the cleavage of peptidoglycan, the major constituent of gram-positive bacterial cell wall [54]. The cutting of even a small number of these polysaccharide chains leads to the cell wall rupture and in turn to the bacteria cell burst as a result of the osmotic stress [55]. Therefore, Lyz-AuNPs complexes represent a promising specimen of nanozymes. In this framework, we analysed the electrostatic features of lysozyme and the SR-microFTIR spectra of Lyz-AuNPs, to investigate the orientation and folding of the proteins adsorbed to AuNPs, directly involved in the functionality of the complexes. The active site of the protein consists of two amino acids, Glu35 and Asp52, which are negatively charged at pH 6.5 when the enzyme is active. This feature is highlighted in the electrostatic potential map of the lysozyme of Figure S13 as compared to the corresponding map at pH 4, where the active site loses the negative charge. The charge distribution at physiological pH suggests an orientation of the proteins adsorbed onto the anionic AuNPs with the active site exposed outwards due to electrostatic repulsion, in accordance with the study of Zhang *et al.* [56] which claims that the binding site of lysozyme with 90 nm AuNPs involves the residues Phe3, Cys6 and Cys127. In addition, this region of the protein contains a consistent number of positive residues [57], promoting the electrostatic attraction of the protein with the citrate ions on the AuNPs surface [16].

The spectroscopic analysis of the lysozyme folding was performed by comparing SR-microFTIR spectra of Lyz-AuNPs and free lysozyme as reported in Section S10 of ESI. The analysis of the Amide I band, based on the deconvolution of the spectral components associated to the different secondary structures of the protein [58], excludes serious denaturation when the protein molecules adsorb to AuNPs.

Proceeding from these considerations, the efficiency of the Lyz-AuNPs complexes as nanozyme was tested by means of a turbidimetric activity assay performed on *Micrococcus lysodeikticus* bacteria. Briefly, we added aliquots of the sample to a bacterial suspension and monitored in time the subsequent decrease in the absorbance at 450 nm due to bacteria death, to derive the activity according to equation 2. In Figure 6A we report a representative example of the obtained absorbance trend, measured on the Lyz-AuNPs sample prepared at $x = 5000$, compared to that of free lysozyme proteins at the same concentration, together with the absorbance trends of the bacterial suspension and bare AuNPs as controls. A significant lysis capability of the Lyz-AuNPs samples in comparison to the bacteria control and to the bare AuNPs can be clearly recognised.

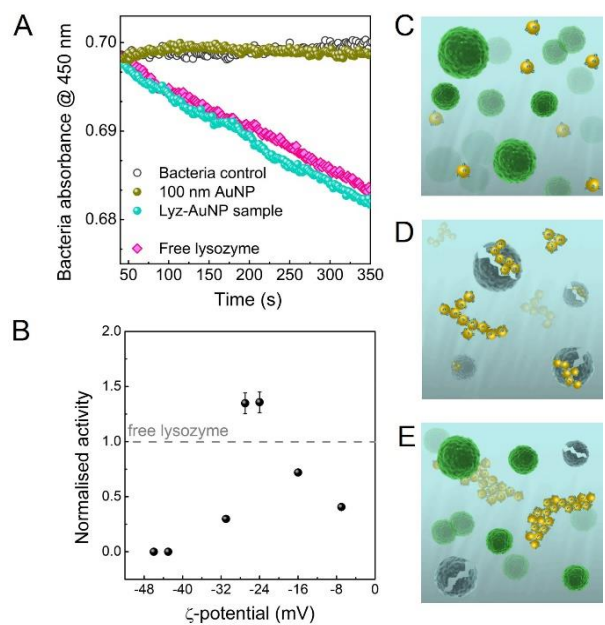

**Figure 6.** Lysozyme activity assay performed on the Lyz-AuNPs complexes made up of 100 nm AuNPs. (A) Representative analysis on the sample prepared at $x = 5000$: absorbance at 450 nm of the bacterial suspension as a function of time for Lyz-AuNPs complexes (light blue points), lysozyme proteins at the same concentration (pink diamonds); controls: bacteria suspension (empty circles) and bare 100 nm AuNPs (dark yellow points). (B) Normalised activity of the Lyz-AuNPs complexes, calculated according to equation 3, as a function of their ζ-potential. Each reported value is obtained by three measurements. Scheme of the hypothesised mechanisms of action in performing antibacterial activity depending on the ζ-potential of Lyz-AuNPs complexes: (C) single Lyz-AuNPs colloids, (D) small aggregates and (E) large aggregates.

Moreover, the steeper slope in the absorbance trend of the Lyz-AuNPs sample with respect to the free protein points out a faster activity rate related to the presence of the complexes.

To better investigate this aspect, we tested the enzymatic activity at varying the lysozyme-AuNP molar ratio. For each sample, we also measured the activity of non-adsorbed proteins, separated from the Lyz-AuNPs complexes by centrifugation, and that of free lysozyme at the same concentration used in the synthesis protocol.

Representative absorbance trends for all the samples are reported in Figure S15 of ESI. In all the analysed samples, the measured activity of non-adsorbed lysozyme resulted lower than both that of the full system and that of the free lysozyme at the same concentration used to prepare the samples. These findings highlight the contribution of Lyz-AuNPs complexes, given by the difference between the activity of the sample and that of the supernatant. It is worth noting here that the concentration of lysozyme employed in the samples is of the order of magnitude of 0.1 μg/mL. This value is markedly low if compared to other antibacterial nanomaterials reported in the literature which use lysozyme as active component. In fact, employing nanoparticles of different materials, the lysozyme concentrations at which significant activity has been observed are of the order of 1 μg/mL in the case of Ag [59], of 10 μg/mL for ZnO and Se [60,61], up to 10 mg/mL in the case of $SiO_2$ [62].

To quantify the antibacterial performances of Lyz-AuNPs in comparison to the free protein, we defined in equation 3 an experimental parameter, namely the normalised activity, as the ratio between the complexes activity and that of the same amount of free lysozyme. The calculated values are reported in Figure 6B as a function of the ζ-potential of the complexes. Noteworthy, it is possible to extrapolate a correspondence between the surface charge of the aggregates and their lysis capability, identifying three different scenarios: *i)* at high absolute values of the ζ-potential the system does not show a significant activity; *ii)* at intermediate values, starting from the aggregation threshold, the efficiency in the lysis capability increases until reaching an enhancement of the activity with respect to free proteins; *iii)* at low values, the catalytic efficiency of the complexes decreases. This behaviour can be interpreted with reference to previous studies which report on the influence of the interfacial electrostatic potential of different bacteria on the interaction with nanoparticles [62-64]. In particular, the gram-positive bacteria employed in this study have a negative ζ-potential of -15.6 mV (see Figure S16 of ESI), consistent with the value reported in literature [65]. More in detail, at low molar ratios the highly negative ζ-potential of the single AuNPs prevents the interaction with bacteria (Figure 6C). At increasing the molar ratio, the electrostatic repulsion between the Lyz-AuNPs and the bacterial wall progressively decreases, and the system shows an increasing lysis capability. In particular, for ζ-potential values in the range between -27 mV and -24 mV, Lyz-AuNP complexes appear more efficient as compared to the free protein. The decrease of the ζ-potential implies that the attractive specific interactions between the enzyme and its substrate progressively overcome the repulsive electrostatic ones, thus promoting the binding of the aggregates on the bacteria membrane. The huge concentration of protein confined onto the AuNPs results in the locally enhanced cleavage, with the hydrolysis of several polysaccharide chains in the same portion of the cell wall. This favours and accelerates the burst of bacteria, leading to a higher catalytic activity (Figure 6D).

When the hydrodynamic diameter of the Lyz-AuNPs complexes rises to several hundred of nanometres their diffusivity decreases remarkably, thus clusters hit the bacteria wall with an extremely lowered rate when compared to the free protein. In addition, a relevant amount of proteins results hindered within clusters and does not have access to the bacteria wall, making the system less efficient in comparison to smaller clusters. This effect could explain the decreasing trend observed in the activity of samples with low absolute values of the ζ-potential (Figure 6E).

The analysis herein reported points out that the antibacterial activity of the complexes can be tuned through their aggregation, witnessing the pivotal role of the controlled assembly provided by patchy interactions. In the perspective of developing an active bio-plasmonic nanozyme system, the thermoplasmonic properties of AuNPs could be exploited to enhance the antimicrobial activity of the Lyz-AuNPs complexes.

## Conclusions

We realised a plasmonic active nanozyme with antibacterial properties based on the controlled aggregation of lysozyme decorated gold nanoparticles (Lyz-AuNPs). We demonstrated the aggregation-dependent modulation of the antibacterial activity of the nanomaterial and highlighted the possibility of acting on the assembly process to reach the tunability of both the optical response and catalytic activity.

In this respect, we exploited the key role of charge-patch interactions in generating self-limiting clustering of the colloids for carrying on a detailed analysis and a comprehensive rationalisation of the synthesis of Lyz-AuNPs assemblies, that allowed us to obtain a dispersion of stable clusters with selectable size. We therefore focused on the antibacterial properties of the nanomaterial in relation to the assembly process. Specifically, the spectroscopic study of the extinction profiles assessed the strict correspondence between the clusters morphology and their plasmonic response, and pointed out the possibility of exploiting plasmon hybridisation to spread the extinction bands up to the near infrared, which is particularly suitable for *in vivo* application due to the high transmission of biological tissues in this spectral range. The catalytic efficiency of the system was investigated on living bacteria, pointing out its aggregation-dependent modulation and highlighting an enhancement of the performances for assemblies of the proper size.

Our integrated study represents an important advancement in the synthesis of nanozymes as compared to currently published literature [13-15], providing pivotal insights on the role of the aggregation in determining the properties of the nanomaterial and its efficiency, and highlighting the possibility of controlling and optimising the features involved in the antibacterial function. We are confident that our findings represent a promising starting point for the development of novel antibacterial nanozymes with highly controllable activity and optical properties for effective application in treating multi drug resistant bacteria.

On the basis of the obtained results, future work will be aimed on one hand at further exploring the basic aspects of the clustering process of plasmonic nanoparticles by a detailed analysis of the morphology in correlation with the optical response to further improve our control on the antibacterial functionality, and on the other hand at investigating the interplay between the different antibacterial functions of Lyz-AuNPs with particular focus on the thermoplasmonic contribution in killing bacteria.

## Acknowledgements

Authors acknowledge the Physics Department of Sapienza University of Rome for providing access to the Sapienza Nanoscience & Nanotechnology Laboratories (SNN-Lab) of the Research Center on Nanotechnology Applied to Engineering of Sapienza University (CNIS) for AFM and SEM measurements. Authors acknowledge SOLEIL for providing synchrotron radiation facilities under proposal n. 20181452 at the SMIS beamline and under proposal n. 20180833 at the SWING beamline. Authors thank Thomas Bizien for assistance during the SAXS experiment. The research leading to SR-microFTIR and SAXS results has been supported by the project CALIPSOplus under the Grant Agreement 730872 from the EU Framework Programme for Research and Innovation HORIZON 2020. A.C. acknowledges Sapienza University of Rome for the grant "Progetti per Avvio alla Ricerca", prot. n. AR11715C821B8F01.

## Conflicts of interest

There are no conflicts to declare.